# Memristive effects in YBa$_2$Cu$_3$O$_{7-x}$ devices with transistor-like structure

Aurélien Lagarrigue, Carolina de Dios, Salvatore Mesoraca, Santiago Carreira, Vincent Humbert, Javier Briatico, Juan Trastoy and Javier E. Villegas[*]

[1]Unité Mixte de Physique, CNRS, Thales, Université Paris-Saclay, 91767, Palaiseau, France

Cuprate superconductors are strongly sensitive materials to disorder and oxygen stoichiometry; even minute variations of those parameters drastically change their electronic properties. Here we exploit this characteristic to engineer a memristive device based on the high-$T_C$ superconductor YBa$_2$Cu$_3$O$_{7-x}$ (YBCO), in which local changes of the oxygen content and induced disorder are exploited to produce memory effects. These effects are triggered electrically in a three-terminal device whose structure is reminiscent of a transistor, consisting of a YBCO channel and an Al gate. The Al/YBCO interface, which controls the gate conductance, displays a giant, bipolar, reversible switching across a continuum of non-volatile conductance states that span over two Decades. This phenomenon is controlled by the gate voltage magnitude and is caused by oxygen exchange between YBCO and Al. Concomitantly, the channel shows a gradual, irreversible superconductor-to-insulator transition that retains a memory of the power dissipated in the device, and can be explained by induced bulk disorder. The observed effects, and the understanding of the interplay between the underlying mechanisms, constitute interesting ingredients for the design and realization of novel memristors and switches for superconducting electronics.

[*]email: javier.villegas@cnrs-thales.fr

**Introduction**

Memristors, named and defined in 1971 by Leon Chua as resistors whose properties depend on their electrical history[1], are multi-state memories[2,3] that play a prominent role in the development of novel neuromorphic computing [4], logic operations [5] as well as signal processing circuits [6,7]. The realization of memristors based on superconducting materials is interesting for various reasons. On the one hand, the development of non-volatile memory in superconducting circuits attracts continued interest [8] as it constitutes a missing, game-changing ingredient in the field of conventional superconducting electronics [9–14]. On the other hand, the implementation of memristive effects in superconducting devices is arguably pivotal for the nascent field of superconducting neuromorphics [15–18], and will expectedly open new avenues in other areas such as quantum computing [19].

Among the superconducting materials in which memory effects could be sought, high-critical temperature ($T_C$) cuprates [20,21] are especially appealing. This is, first and most obviously, because devices based on them can operate at temperatures in the tens of K range, dramatically reducing the burden of cryogenics as compared to low-$T_C$ superconductors. This has indeed fostered a variety of electronic applications and technologies [9–14,22,23]. Second, and importantly, cuprates belong to the vast family of strongly-correlated functional oxides [24] in which the presence of oxygen vacancies and disorder dramatically changes the ground state. Beyond its fundamental interest, this characteristic offers possibilities for engineering interesting functionalities, as we illustrate with the current experiments.

While the microscopic mechanism for high-$T_C$ superconductivity in the cuprates remains a long-standing problem, it is well established that the CuO$_2$ planes characteristic of their crystal structure play a significant role. Indeed, the superconducting properties are strongly sensitive to the presence of disorder as well as the oxygen stoichiometry in such CuO$_2$ planes [25–27]. For example, the oxygen stoichiometry controls the doping level, leading to a complex temperature-doping phase diagram which includes the superconducting and insulator phases, as well as other exotic ones, such as the pseudo-gap [28] or the strange metal phase [29]. An interesting consequence of the intricate phase diagram and its connection to oxygen content and disorder is the possibility of tuning the ground state through the controlled migration of ions. Such control can be achieved electrically, via the application of currents or voltages. One of the possible oxygen-control approaches consists of circulating large current densities, as reported in thin films and nanoconstrictions[30–35]. It was shown in these experiments that a sufficiently large current (typically above $7\ MA \cdot cm^{-2}$) [33–36] can produce direct electromigration and, under certain conditions, the local heating produced by lower current densities can lead to the thermomigration of oxygen ions [34,37]. Both current-driven mechanisms can change the cuprate's doping level and physical properties at a local scale [38,39]. A different approach for electrically controlling a cuprate's ground state consists of placing a low-reduction-potential material in contact with YBCO to exploit the redox reaction that occurs at the interface between both materials. The oxygen exchange with the second material can be tuned by applying voltage pulses of different polarity and amplitude, which drive oxygen in and out of the cuprate at will, thus allowing for a reversible switching of the transport and superconducting properties [40–43].

In this work, we study devices with a structure analogous to that of a transistor, in which the channel is made of the archetypal high-critical temperature superconductor $YBa_2Cu_3O_{7-\delta}$ (YBCO) and the gate is made of Al (see **Fig. 1**). Interestingly, we show that both redox and current-induced mechanisms play a role in these devices. On the one hand, the redox mechanism allows for reversibly switching the gate conductance via changes in the oxidation state at the Al/YBCO interface. In particular, the gate conductance shows a memristive behaviour: it can be switched across a continuum of non-volatile levels depending on the history of applied gate pulses. Those levels are intermediate between a high-conductance (hereafter ON) state and a low-conductance (OFF) state, the ratio between the ON/OFF conductance being as high as 10000 %. This memristive behaviour is observed within a range of low to moderate gate voltages (depending on the device, this range can be as high as $\sim 15\ V$). On the other hand, the gate voltage also produces a non-reversible tuning of the YBCO channel's ground state, namely a superconductor-to-insulator transition, which is caused by ion migration within the bulk of the YBCO film that changes the source-drain resistance. This effect becomes stronger at higher gate voltages, for which the electrical power dissipated in the channel is significantly higher and introduces bulk disorder via thermal effects (structural damage caused by local heating, likely ion thermomigration). The obtained results and understanding pave the way for devices in which the dominance and/or interplay of redox and dissipation effects can be controlled to optimize specific applications −for example superconducting memristive or current-limiting elements in superconducting circuits− and contribute to enlarging the technological prospects of cuprate superconductors.

**Method**

The structure of the studied devices is shown schematically in **Fig. 1(c)**, and presents 3 terminals, named by analogy with a conventional transistor: gate (G), drain (D) and source (S). The device fabrication is based on commercial (Ceraco ceramic coating GmbH) 150 $nm$-thick c-axis YBCO films grown on r-cut sapphire substrates, with a buffer layer of cerium oxide ($CeO_2$) and capped *in situ* with 100 $nm$ of gold. Those films were patterned using various steps of optical lithography and Ar ion-beam etching. The first step is to define the outline of the devices, in which the YBCO/Au bilayer is etched down to the $CeO_2$ layer [**Fig. 1(a)**]. Then, the thickness of the superconducting channel $t_{YBCO}$ was defined by local etching [**Fig. 1(b)**]. This process was monitored using a secondary ion mass spectrometer (SIMS). The sample topography was subsequently studied with atomic force microscopy (AFM), which allowed us to precisely determine the YBCO thicknesses $t_{YBCO}$. We fabricated different samples with 35 nm < $t_{YBCO}$ < 140 nm. Reducing $t_{YBCO}$ from 150 $nm$ to 140 $nm$ yields $T_C \approx 86\ K$. Further thinning to $t_{YBCO} = 35\ nm$ lowers the $T_C$ to ~72 $K$, which is likely due to moderate oxygen loss caused by heating during the ion etching. Subsequent optical lithography, plasma cleaning, sputtering deposition and lift-off steps allow defining the gate (denoted G), made of aluminium ($t_{Al} = 2\ nm$ or $t_{Al} = 10\ nm$) covered by 100 $nm$ of gold [**Fig. 1(c)**]. The gate width $W_{Al}$ was either 60 µm or 5 µm, depending on the device, and thus the interface's area was either $60 \times 100\ \mu m^2$ or $5 \times 100\ \mu m^2$. In addition to the main devices, in which Al is in direct contact with YBCO under the gate, we fabricated two control samples whose gates are made solely of Au.

To electrically characterize the devices, we measured the gate-drain conductance as well as probed the source-drain channel resistance after applying gate voltages, always in a two-probe configuration. For the former type of measurement, the gate-drain was voltage-biased and the current was probed using a Keithley 2450 source-meter [upper sketch in **Fig. 1(d)**]. For the latter, a Keysight B2901B precision current source and a Keithley 2182A nanovoltmeter were used for the current biasing and voltage measurements, respectively [lower sketch in **Fig.1(d)**].

**Results**

*Gate-drain measurements*

A first set of measurements demonstrating the switching behaviour of the gate is shown in **Fig. 2**. In these measurements, the gate voltage $V_G$ is ramped from zero to a final value $V_{write}$ (which can be positive or negative) and then back to zero in a fixed time (typically 20 seconds) using a large number of steps (typically 100), as illustrated by the inset of **Fig. 2(a)**. The gate-drain current is measured during the $V_G$ sweep to obtain $I_G(V_G)$ characteristics as those displayed in the main panel of **Fig. 2(a)**. Typically, a data set contains $I_G(V_G)$ curves for forty or eighty different $V_{write}$, although the figure only displays a selected subset for clarity. Notice that $V_{write}$ is varied in a particular chronological sequence, which is indicated in the colour legend (top to bottom). Direct inspection of **Fig. 2(a)** reveals a jump between a low-conducting (OFF) state to a high-conducting (ON) one as $V_{write}$ is increased above a positive threshold (~4.5 - 5 V), and the opposite (yet more gradual) switching (ON to OFF) as $V_{write}$ is decreased below ~ -1.5 V. To better describe that behaviour, we have selected the $I_G(V_G)$ curve set in the six different panels of **Fig. 2(b)**. Each panel corresponds to a different regime. Starting from the OFF state, the first regime [**Fig. 2(b), 1**] corresponds to positive voltage excursions with $0 < V_{write} < 4\,V$. Here the I-V curves are reversible and overlap, that is, the gate's conducting state is unchanged by the voltage excursions. The second regime appears as $V_{write}$ exceeds ~4.5 V [**Fig. 2(b), 2**]: here the $I_G(V_G)$ becomes more and more hysteretic as $V_{write}$ is increased, the conductance being the highest in the voltage-decreasing branch. Near $V_{write}$ ~ 5 V the hysteresis becomes the largest, indicating the full switching from the OFF into the highest conductance ON state. This state remains stable both if $V_{write}$ is further increased, and also in regimes 3 [**Fig. 2(b), 3**; $5 > V_{write} > 0\,V$] and 4 [**Fig. 2(b), 4**; $0 > V_{write} > -1.5\,V$]. Regime 5 [**Fig. 2(b), 5**] corresponds to the gradual switching from the ON into the OFF state, for voltage excursions with $V_{write}$ below -1.5 V. Once $V_{write}$ ~ $-5\,V$ is reached, the OFF state is fully set and we enter the 6$^{th}$ regime [**Fig. 2(b), 6**], which corresponds to a stable OFF state as $V_{write}$ is increased from $-5\,V$ to $0\,V$.

Additionally, the above memristive behaviour and the possibility to stabilize intermediate states between ON and OFF, can be easily sensed via low-bias conductance measurements. That is demonstrated in **Fig. 3(a)** for which we applied $V_{write}$ voltages following the sequence shown in the legend (bottom to top), for a different device compared to **Fig. 2** (see the captions). After each $V_{write}$, we measured a low-bias $I_G(V_G)$ of the gate with $|V_G| < 0.5\,V$, which is low enough to probe the remnant state of the gate without modifying it. This allow us to numerically calculate the differential conductance $G = dI_G/dV_G$ as a function of the bias. One can also observe the gradual switching between the OFF and ON states across intermediate ones, as $V_{write}$ is increased from $4.7\,V$ to $7\,V$. **Fig. 3(b)** shows the differential conductance for a fixed $V_G = 100\,mV$ as a function of $V_{write}$, which is cycled from positive to negative to drive the ON/OFF conductance switching. The behaviour is typical of all the samples with Al gating:

a hysteretic, reversible ON/OFF switching, with the conductance change being as large as 10000 % (in the example, from $10^{-6}\ S$ to $10^{-8}\ S$). Notice also the strong asymmetry in the voltage required for initiating the switching from OFF to ON $|V_{POL}(+)| = 4.7\ V$ and *vice versa* $|V_{POL}(-)| = 0.7$ V. **Fig. 3(c)** shows the typical behaviour of the control samples, which have a noble-metal gate (Au instead of Al). Their behaviour shows three main differences as compared to the Al devices. First, the conductance level is overall three to five orders of magnitude higher for Au than for Al. Second, for Au, the conductance switching is extremely small –only ~10 % as opposed to ~10000 %. This suggests that a material with a tendency to oxidize (such as Al) is required to produce large conductance-switching effects. Second, with the Au gate the $G(V_{write})$ cycle is fully symmetric: $|V_{POL}(+)| = |V_{POL}(-)|$. Indeed, as it will be discussed later, the asymmetry of the conductance cycles characteristic of the devices with Al gate is a fingerprint of the redox reaction underlying the large resistance switching [43].

*Source-drain measurements*

We measured the source-drain conductance after applying $V_{write}$ to determine its impact on the YBCO channel. **Fig. 4(a)** shows the channel's resistance $R_{SD}$ as a function of the temperature, measured after applying increasing $V_{write}$ (see legend). The measurements shown in **Fig. 4(a)** are representative of all the devices with an Al gate. One can distinguish two main features. First, increasing $V_{write}$ leads to a gradual increase of the channel's normal-state resistance (for $T > \sim 90\ K$) until, ultimately, for $V_{write} = 15\ V$ a highly resistive state is observed at all temperatures. Second, the normal-to-superconducting resistive transition, apparent for $V_{write} \leq 12\ V$, is gradually enlarged as $V_{write}$ increases. Notice that the superconducting transition onset occurs at a similar temperature for all $V_{write} \leq 12\ V$, as it only decreases from $T_C \sim 88$ K for the pristine YBCO channel to $T_C \sim 86$ K for $V_{write} = 12\ V$. **Fig. 4(b)** shows similar measurements carried out on a control sample whose gate is made of gold instead of Al. These measurements globally show the same behaviour as the one described for the Al-gated device: as $V_{write}$ is increased, the normal-state resistance increases and the superconducting transition widens. To provide a more quantitative analysis of the latter observation, we define in our measurements the effective critical temperature $T_C^*$ as the temperature for which the channel resistance drops below 1 Ω (this roughly corresponds to the residual resistance due to the 2-probe configuration). This criterion is shown by the black dash-dotted horizontal line displayed in **Fig. 4(a)** and **(b)**. **Fig. 4(c)** shows $T_C^*$ as a function of $V_{write}$ for different devices with a channel thickness varying between $50\ nm < t_{YBCO} < 125\ nm$. In all cases, we observe a monotonic decrease of $T_C^*$ with increasing $V_{write}$, which allows driving the channel from superconducting into a highly resistive state. The voltage required to suppress superconductivity scales with the thickness of the YBCO channel: the thicker the channel, the higher the required $V_{write}$. Interestingly, a similar suppression of superconductivity is observed for the control samples whose gate is made of Au [see **Fig. 4(b)** and the corresponding purple curve in **Fig. 4(c)**]. That is, contrary to the conductance switching effects discussed in **Fig. 3**, here similar effects are observed regardless of whether the gate material can reduce YBCO or not. Another key observation is that, at variance to the gate conductance-switching effects, the modulation of superconductivity in the YBCO channel is irreversible: the application of negative $V_{write}$ voltage pulses cannot enhance $T_C^*$ nor drive a decrease in the channel's resistance. This is well illustrated in **Fig. 5**, which monitors the concomitant gate and channel resistance changes induced by the application of a series of $V_{write}$ cycles. **Fig. 5(a)** shows the sequence of $V_{write}$, which goes back and forth from positive to negative with increasing amplitudes. **Fig. 5(b)** shows the remnant channel's resistance $R_{SD} = V_{SD}/I_{SD}$ for $I_{SD} = 1$ µA, (black curve) and the gate's resistance $R_G = V_G/I_G$ for $I_G = 1$ µA (green curve), measured following each writing cycle. One can see that the gate's resistance $R_G$ is reversibly switching between two ON and OFF levels, which are roughly stable after the 1st cycle, regardless of the gradual increase of $V_{write}$ across the sequence. Contrarily, the channel displays irreversible resistance-increasing steps, one switching cycle after the other. The unavoidable conclusion is that the gate and channel resistance variations arise from different mechanisms.

An important key to understanding the irreversible channel's resistance variation across the sequence of $V_{write}$ is given in **Fig. 5(c)**. This figure displays (red curve) the maximum power dissipated in the device $P_{max} = V_{write} \cdot I(V_{write})$ as the gate voltage is cycled to $V_{write}$. Direct comparison between the red curve in **Fig. 5(c)** and the black one in **Fig. 5(b)** demonstrates that the channel's resistance $R_{SD}$ is a step function of the dissipated power. Interestingly, the threshold power $P_{th}$ at which the channel's resistance increase is triggered is not constant, but rapidly increases during the first $V_{write}$ cycles and then tends to saturate, as shown by the blue curve in **Fig. 5(c)**. This means that the process that leads to the suppression of the channel's conducting (and superconducting) properties is self-limiting: once a given power threshold triggers the increase of the channel's resistance, this will not further change unless a higher power is applied. Interestingly, this allows reaching a regime, after a few $V_{write}$ cycles (training), in which the memristive behaviour of the gate [green curve in **Fig. 5(b)**] can be exploited without inducing further changes in the channel. Conversely, if desired, the channel can be driven to a lower conductance state by increasing $V_{write}$, while the memristive behaviour of the gate is preserved. As demonstrated in **Fig. 4(c)** and further discussed below, the onset of the different regimes can be engineered by changing the thickness of the YBCO channel.

**Discussion**

The reversible switching of the conductance between ON, OFF and intermediate non-volatile states [**Fig. 3**] can be explained by an interfacial redox reaction between aluminium and YBCO, similarly as for the memristive effects observed in YBCO/MoSi [43] and NdNiO₃/MoSi [44] micro junctions. The redox reaction naturally reduces the material with a higher reduction potential (here YBCO, given the high reduction potential of copper $E°(Cu) = 2.4 \, eV$) and oxidizes the material with a lower reduction potential ($E°(Al) = -1.676 \, eV$)[45]. As a result, a 2-3 $nm$ thick oxygen-depleted YBCO layer is expected at one side of the interface [43], and conversely, an AlO$_x$ layer will form at the other side. Because both oxygen-depleted YBCO and AlO$_x$ are insulating [46–48], a tunnel barrier forms at the interface, which strongly diminishes its electrical conductance. This explains the low-conductance OFF state. That redox reaction can be reversed by the application of a $V_{write}$ that exceeds the difference between the reduction potentials, very much as a battery can be charged, leading to oxygen transfer from AlO$_x$ into YBCO, and thus to a thinning of the oxygen-depleted YBCO and AlO$_x$ tunnelling barrier. This explains the eventual switching into the highest-conductance ON state upon application of a sufficiently high positive $V_{write}$. This redox scenario is supported by various key observations. As shown by the examples displayed in **Fig. 3(a)**, the conductance shows a non-ohmic behaviour similar to that observed earlier in YBCO/MoSi junctions [43], which is consistent with electron tunnelling across the Al/YBCO interface. The experimental values of $V_{POL}$ are consistent with the expectations [43,44]. In particular, we find that the voltage $V_{POL}(+)$ required for switching into the ON state is around $4.7 \, V$, which is above the minimum $\Delta E = E°(Cu) - E°(Al) = 2.4 - (-1.676) = 4.076 \, eV$ [45] expected from electrochemical arguments. The excess voltage of nearly $\sim 0.7 \, V$ is required to overcome the barrier for oxygen ion diffusion, that is, to accelerate the dynamics of the electrochemical reaction. This is consistent with $V_{POL}(-) \approx -0.7V$ required to switch back into the OFF state, which is spontaneous from the electrochemical point of view and thus only requires activation over the barrier for oxygen diffusion. Therefore, the redox scenario naturally explains the asymmetry in the switching loop of **Fig. 3(b)**. This scenario is also supported by the behaviour of the control samples whose gate is made of Au instead of Al. These show three to four orders-of-magnitude larger conductance, and hundredfold weaker conductance-switching effects. That is as expected from the redox scenario, which cannot occur at the interface between YBCO and a noble metal. The small conductance switching observed in **Fig. 3(c)** for Au can be explained by changes in the distribution of the fewer, native oxygen vacancies in YBCO, which are attracted towards/repelled from the interface under the influence of the local electric field induced by $V_{write}$[43]. This is consistent with the fact that, in the case of the Au gate, the switching loop [**Fig. 3(c)**] is fully hysteretic with $|V_{POL}(+)| = |V_{POL}(-)| \approx 0.7\ V$. As discussed above, this is identical to the voltage required to switch from the ON to OFF state in the samples with the Al gate, which further substantiates that 0.7 V is the voltage necessary to overcome the barrier for oxygen ion diffusion.

Because the changes in the YBCO channel resistance are irreversible, similar in devices with Al and Au gates [**Fig.4** and **Fig.5**], and ultimately triggered by the dissipated power [**Fig. 5(c)**], we conclude that they are not caused by the interfacial redox reaction. Having ruled out this mechanism, we turn to current-induced effects. The normal-state resistivity increase and $T_C$ depression indicates either that the oxygen content of the channel is reduced (underdoping) [35,49], that disorder is being introduced [50], or both. The current densities circulating across the device upon application of $V_{write}$ range between $1\ kA \cdot cm^{-2}$ and $10\ kA \cdot cm^{-2}$, based on the measured currents and considering the section of the YBCO channel. However, these current densities are well below the value typically required to produce electromigration in YBCO ($\sim 7\ MA \cdot cm^{-2}$) [33–36]. Consequently, it is unlikely that oxygen displacement due to electron collisions may explain the resistance increase and $T_C$ depression observed in **Fig. 4**. Furthermore, electromigration is expectedly reversible [35], meaning that reversing the polarity of the current should allow moving oxygen back and recovering both $T_C$ and resistivity. This is not observed. Altogether, the experimental observations rule out that electrochemistry and electromigration may contribute significantly to the changes in the channel's conducting and superconducting properties. Instead, the fact that the resistance enhancement is triggered by power dissipation [**Fig. 5(c)**] suggests that the observed effects are linked to Joule heating. A related mechanism is thermomigration [51]: the heating produced by local dissipation creates a temperature gradient across the device, which leads to migration of the most mobile ions. This mechanism has been discussed earlier in experiments with YBCO nanowires [34,37]. In these, thermomigration of the highly mobile oxygen $O^{-2}$ in the $CuO_2$ planes leads to underdoping. In our experiments, this power dissipation seems to be producing an inhomogeneous effect. Should thermomigration lead to homogeneous underdoping of the YBCO channel in our devices, we would observe a depression of $T_C$ with no significant widening of the transition, as well as a characteristic departure from the linear temperature dependence of the normal-state resistivity [35,52]. In contrast, **Fig. 4(a)** and **(b)** show a widening of the superconducting transition with an onset temperature that remains essentially unchanged ($86\,K - 88\,K$) for increasing $V_{write}$. Furthermore, despite the strong overall enhancement of the normal-state resistivity, its linear temperature dependence is preserved for all $V_{write}$, except the highest for which the channel becomes insulating (figure not shown). All of that indicates that in our experiments power dissipation is not producing a homogeneous underdoping of the YBCO channel. Instead, the experimental observations consistently point to inhomogeneous changes within the channel, and to a dominant role of heat-induced local structural disorder which increses resistivity and depresses $T_C$ due to pair-breaking by enhanced eletronic scattering, to which d-wave superconductors are extremely sensitive [50,53]. In particular, a consistent picture sketched in **Fig. 1(d)** is that bulk structural disorder is created locally under the gate, where YBCO is thinner, over a volume that extends deeper after each writing cycle. This explains the observation of an almost constant superconduting transition temperature onset (that corresponds to the YBCO volume further from the gate) despite the large resistance increase (created by the the growing volume of disordered YBCO) as well as to the large widening of the superonducting transition.

**Conclusions**

We have realized three-terminal superconducting devices with a transistor-like structure based on YBCO and Al. The Al/YBCO gate interface shows memristive behaviour: a large, reversible switching of the conductance across a continuum of non-volatile states driven by the application of gate voltages, which is produced by a reversible redox reaction. On the other hand, the channel's resistance and superconducting $T_C$ are modified by the power dissipated in the device, which depresses the superconducting properties due to induced disorder. Interestingly, the latter process is self-limiting and, after a few training cycles, allows exploiting the memristive function without further modification of the channel's properties. Conversely, by increasing the dissipated power, one can tune the properties of the channel without affecting the memristive functions. The latter could be exploited, for example, for current-limiting functions intended at protecting superconducting circuitry, or for detection of (cumulative) power dissipation. Those functions could be applied in areas like superconducting signal processing [6,7], where memory and switching functions are required but hard to implement [8], or in the nascent area of neuromorphic computing with superconductors [18]. The technological prospects in these areas are enhanced by the fact that the present demonstration is based on high-temperature superconductors. Furthermore, we believe that the identification of the different involved mechanisms we provide will be beneficial for the design of other cuprate-based (or more generally, complex-oxide-based) electronic devices subject to moderate to high voltages and power dissipation. Finally, it is worth mentioning that the observation of large redox-induced tunnel-conductance switching was somewhat unexpected because the gate areas here are orders of magnitude bigger than those of the micro junction where they have been demonstrated earlier[43]. Indeed, one would have anticipated that, in such large junctions, the likely presence of defects would have prevented the observation of the very large conductance switching effects. Their present realization shows that the redox mechanism is very robust against defects and can be easily implemented at a larger scale, which can be relevant for specific applications.

# Acknowledgements

Work supported by ERC PoC 966735 "SUPERMEM", French ANR-17-CE30-0018-04 "OPTOFLUXONICS", Flag-ERA "To2DoX" and Cost Action 21144 "SUPERQUMAP". A.L. acknowledges the support from the French ANRT in the form of a CIFRE scholarship.

[1]     Chua L O 1971 Memristor—The Missing Circuit Element *IEEE Trans. Circuit Theory* **18** 507–19

[2]     Ho Y, Huang G M and Li P 2009 Nonvolatile memristor memory: Device characteristics and design implications *IEEE/ACM Int. Conf. Comput. Des. Dig. Tech. Pap. ICCAD* 485–90

[3]     Parajuli S, Budhathoki R K and Kim H 2020 Nonvolatile Memory Cell Based on Memristor *Univers. J. Electr. Electron. Eng.* **7** 110–7

[4]     Zhang H T, Panda P, Lin J, Kalcheim Y, Wang K, Freeland J W, Fong D D, Priya S, Schuller I K, Sankaranarayanan S K R S, Roy K and Ramanathan S 2020 Organismic materials for beyond von Neumann machines *Appl. Phys. Rev.* **7** 011309

[5]     Gao C, Li T, Wang T and Cao X 2020 Memristor-Based Logic Gate Circuit *2020 IEEE 3rd Int. Conf. Comput. Commun. Eng. Technol. CCET 2020* 330–3

[6]     Zhong Y, Tang J, Li X, Liang X, Liu Z, Li Y, Xi Y, Yao P, Hao Z, Gao B, Qian H and Wu H 2022 A memristor-based analogue reservoir computing system for real-time and power-efficient signal processing *Nat. Electron.* **5** 672–81

[7]     Mouttet B 2009 Proposal for memristors in signal processing *Lect. Notes Inst. Comput. Sci. Soc. Telecommun. Eng.* **3 LNICST** 11–3

[8]     Hilgenkamp H 2021 Josephson Memories *J. Supercond. Nov. Magn.* **34** 1621–5

[9]     Anders S, Blamire M G, Buchholz F I, Crété D G, Cristiano R, Febvre P, Fritzsch L, Herr A, Il'Ichev E, Kohlmann J, Kunert J, Meyer H G, Niemeyer J, Ortlepp T, Rogalla H, Schurig T, Siegel M, Stolz R, Tarte E, Ter Brake H J M, Toepfer H, Villegier J C, Zagoskin A M and Zorin A B 2010 European roadmap on superconductive electronics – status and perspectives *Phys. C Supercond.* **470** 2079–126

[10]    Koelle D, Ludwig F, Bundesanstalt P, Cryosensors S, Berlin D-, Dantsker E and Clarke J 1999 High-transition-temperature superconducting quantum interference devices *Rev. Mod. Phys.* **71** 631–86

[11]    Cybart S A, Cho E Y, Wong T J, Wehlin B H, Ma M K, Huynh C and Dynes R C 2015 Nano Josephson superconducting tunnel junctions in YBa 2 Cu 3 O 7 – δ directly patterned with a focused helium ion beam *Nat. Nanotechnol.* **10** 598–602

[12]    Bauch T, Lindström T, Tafuri F, Rotoli G, Delsing P, Claeson T and Lombardi F 2006 Quantum dynamics of a d-wave Josephson junction *Science* **311** 57–60

[13]    Ortlepp T, Ariando, Mielke O, Verwijs C J M, Foo K F K, Rogalla H, Uhlmann F H and Hilgenkamp H 2006 Flip-flopping fractional flux quanta *Science (80-. ).* **312** 1495–7

[14]    Ouanani S, Kermorvant J, Ulysse C, Malnou M, Lemaître Y, Marcilhac B, Feuillet-Palma C, Bergeal N, Crété D and Lesueur J 2016 High-Tc superconducting quantum interference filters (SQIFs) made by ion irradiation *Supercond. Sci. Technol.* **29** 094002

[15]    Shainline J M, Buckley S M, Mirin R P and Nam S W 2017 Superconducting Optoelectronic Circuits for Neuromorphic Computing *Phys. Rev. Appl.* **7** 034013

[16]    Cheng R, Goteti U S and Hamilton M C 2019 Superconducting neuromorphic computing using quantum phase-slip junctions *IEEE Trans. Appl. Supercond.* **29** 1–5

[17] Zhang Z, Sun Y and Zhang H T 2022 Quantum nickelate platform for future multidisciplinary research *J. Appl. Phys.* **131** 120901

[18] Schneider M, Toomey E, Rowlands G, Shainline J, Tschirhart P and Segall K 2022 SuperMind: A survey of the potential of superconducting electronics for neuromorphic computing *Supercond. Sci. Technol.* **35** 053001

[19] Salmilehto J, Deppe F, Di Ventra M, Sanz M and Solano E 2017 Quantum memristors with superconducting circuits *Sci. Rep.* **7** 42044

[20] Bednorz J G and Müller K A 1986 Possible high Tc superconductivity in the Ba-La-Cu-O system *Zeitschrift für Phys. B Condens. Matter* **64** 189–93

[21] Proust C and Taillefer L 2019 The remarkable underlying ground states of cuprate superconductors *Annu. Rev. Condens. Matter Phys.* **10** 409–29

[22] Bai D D, Du J, Zhang T and He Y S 2013 A compact high temperature superconducting bandpass filter for integration with a Josephson mixer *J. Appl. Phys.* **114** 133906

[23] Faley M I, Poppe U, Dunin-Borkowski R E, Schiek M, Boers F, Chocholacs H, Dammers J, Eich E, Shah N J, Ermakov A B, Slobodchikov V Y, Maslennikov Y V. and Koshelets V P 2013 High-Tc DC SQUIDs for magnetoencephalography *IEEE Trans. Appl. Supercond.* **23** 1600705–1600705

[24] Bibes M, Villegas J E and Barthélémy A 2011 Ultrathin oxide films and interfaces for electronics and spintronics *Adv. Phys.* **60** 5–84

[25] Matsunaka D, Rodulfo E T and Kasai H 2005 Effects of oxygen-vacancies in high-temperature cuprate superconductors *Solid State Commun.* **134** 355–60

[26] Lee P A, Nagaosa N and Wen X G 2006 Doping a Mott insulator: Physics of high-temperature superconductivity *Rev. Mod. Phys.* **78** 17–85

[27] Chu C W, Deng L Z and Lv B 2015 Hole-doped cuprate high temperature superconductors *Phys. C Supercond. its Appl.* **514** 290–313

[28] Timusk T and Statt B 1999 The pseudogap in high-temperature superconductors: An experimental survey *Reports Prog. Phys.* **62** 61–122

[29] Varma C M 2020 Colloquium: Linear in temperature resistivity and associated mysteries including high temperature superconductivity *Rev. Mod. Phys.* **92** 031001

[30] Vitta S, Stan M A, Warner J D and Alterovitz S A 1991 Electromigration failure in YBa2Cu3O7-x thin films *Appl. Phys. Lett.* **58** 759–61

[31] Moeckly B H, Lathrop D K and Buhrman R A 1993 Electromigration study of oxygen disorder and grain-boundary effects in YBa2Cu3O7- thin films *Phys. Rev. B* **47** 400–17

[32] Moeckly B H, Buhrman R A and Sulewski P E 1994 Micro-Raman spectroscopy of electromigration-induced oxygen vacancy aggregation in YBa2Cu3O7-δ *Appl. Phys. Lett.* **64** 1427–9

[33] Baumans X D A, Fernández-Rodríguez A, Mestres N, Collienne S, Van de Vondel J, Palau A and Silhanek A V. 2019 Electromigration in the dissipative state of high-temperature superconducting bridges *Appl. Phys. Lett.* **114** 012601

[34]   Marinković S, Fernández-Rodríguez A, Collienne S, Alvarez S B, Melinte S, Maiorov B, Rius G, Granados X, Mestres N, Palau A and Silhanek A V. 2020 Direct Visualization of Current-Stimulated Oxygen Migration in YBa2Cu3O7-δThin Films *ACS Nano* **14** 11765–74

[35]   Trabaldo E, Kalaboukhov A, Arpaia R, Wahlberg E, Lombardi F and Bauch T 2022 Mapping the Phase Diagram of a YBa2Cu3 O7-δ Nanowire Through Electromigration *Phys. Rev. Appl.* **17** 024021

[36]   Trabaldo E, Garibaldi A, Lombardi F and Bauch T 2021 Electromigration tuning of the voltage modulation depth in YBa2Cu3O7-δnanowire-based SQUIDs *Supercond. Sci. Technol.* **34** 104001

[37]   Marinković S, Trabaldo E, Collienne S, Lombardi F, Bauch T and Silhanek A V. 2023 Oxygen ordering in untwinned YBa2Cu3 O7-δ films driven by electrothermal stress *Phys. Rev. B* **107** 14208

[38]   Strauven H, Locquet J P, Verbeke O B and Bruynseraede Y 1988 Oxygen evolution from YBa2Cu3O6.85 high Tc superconductors *Solid State Commun.* **65** 293–6

[39]   Bruynseraede Y, Locquet J P, Schuller I K and Van Haesendonck C 1989 Oxygen Disorder Effects in High Tc Superconductors *Phys. Scr.* **29** 100–5

[40]   Leng X, Garcia-Barriocanal J, Bose S, Lee Y and Goldman A M 2011 Electrostatic control of the evolution from a superconducting phase to an insulating phase in ultrathin YBa2Cu3O7-x films *Phys. Rev. Lett.* **107** 6–9

[41]   Palau A, Fernandez-Rodriguez A, Gonzalez-Rosillo J C, Granados X, Coll M, Bozzo B, Ortega-Hernandez R, Suñé J, Mestres N, Obradors X and Puig T 2018 Electrochemical Tuning of Metal Insulator Transition and Nonvolatile Resistive Switching in Superconducting Films *ACS Appl. Mater. Interfaces* **10** 30522–31

[42]   Murray P D, Gilbert D A, Grutter A J, Kirby B J, Hernández-Maldonado D, Varela M, Brubaker Z E, Liyanage W L N C, Chopdekar R V., Taufour V, Zieve R J, Jeffries J R, Arenholz E, Takamura Y, Borchers J A and Liu K 2020 Interfacial-Redox-Induced Tuning of Superconductivity in YBa2Cu3O7-Í´ *ACS Appl. Mater. Interfaces* **12** 4741–8

[43]   Rouco V, Hage R El, Sander A, Grandal J, Seurre K, Palermo X, Briatico J, Collin S, Trastoy J, Bouzehouane K, Buzdin A I, Singh G, Bergeal N, Feuillet-Palma C, Lesueur J, Leon C, Varela M, Santamaría J and Villegas J E 2020 Quasiparticle tunnel electroresistance in superconducting junctions *Nat. Commun.* **11** 658

[44]   Humbert V, El Hage R, Krieger G, Sanchez-Santolino G, Sander A, Collin S, Trastoy J, Briatico J, Santamaria J, Preziosi D and Villegas J E 2022 An Oxygen Vacancy Memristor Ruled by Electron Correlations *Adv. Sci.* **9** 1–10

[45]   P D 1992 CRC Handbook of Chemistry and Physics *J. Mol. Struct.* **268** 320

[46]   Khoobiar S, Carter J L and Lucchesi P J 1968 The electronic properties of aluminum oxide and the chemisorption of water, hydrogen, and oxygen *J. Phys. Chem.* **72** 1682–8

[47]   Abyzov A M 2019 Aluminum Oxide and Alumina Ceramics (review). Part 1. Properties of Al2O3 and Commercial Production of Dispersed Al2O3 *Refract. Ind. Ceram.* **60** 24–32

[48]   Lee P A, Nagaosa N and Wen X G 2006 Doping a Mott insulator: Physics of high-temperature superconductivity *Rev. Mod. Phys.* **78**

[49]   Xi X X, Doughty C, Walkenhorst A, Kwon C, Li Q and Venkatesan T 1992 Effects of field-induced hole-density modulation on normal-state and superconducting transport in YBa2Cu3O7-x *Phys. Rev. Lett.* **68** 1240–3

[50]   Swiecicki I, Ulysse C, Wolf T, Bernard R, Bergeal N, Briatico J, Faini G, Lesueur J and Villegas J E 2012 Strong field-matching effects in superconducting YBa 2Cu 3O 7-δ films with vortex energy landscapes engineered via masked ion irradiation *Phys. Rev. B - Condens. Matter Mater. Phys.* **85** 224502

[51]   Oriani R A 1969 Thermomigration in solid metals *J. Phys. Chem. Solids* **30** 339–51

[52]   Arpaia R, Andersson E, Trabaldo E, Bauch T and Lombardi F 2018 Probing the phase diagram of cuprates with YBa2Cu3 O7-δ thin films and nanowires *Phys. Rev. Mater.* **2**

[53]   Lesueur J, Nedellec P, Bernas H, Burger J P and Dumoulin L 1990 Depairing-like variation of Tc in YBa2Cu3O7-δ *Phys. C Supercond. its Appl.* **167** 1–5

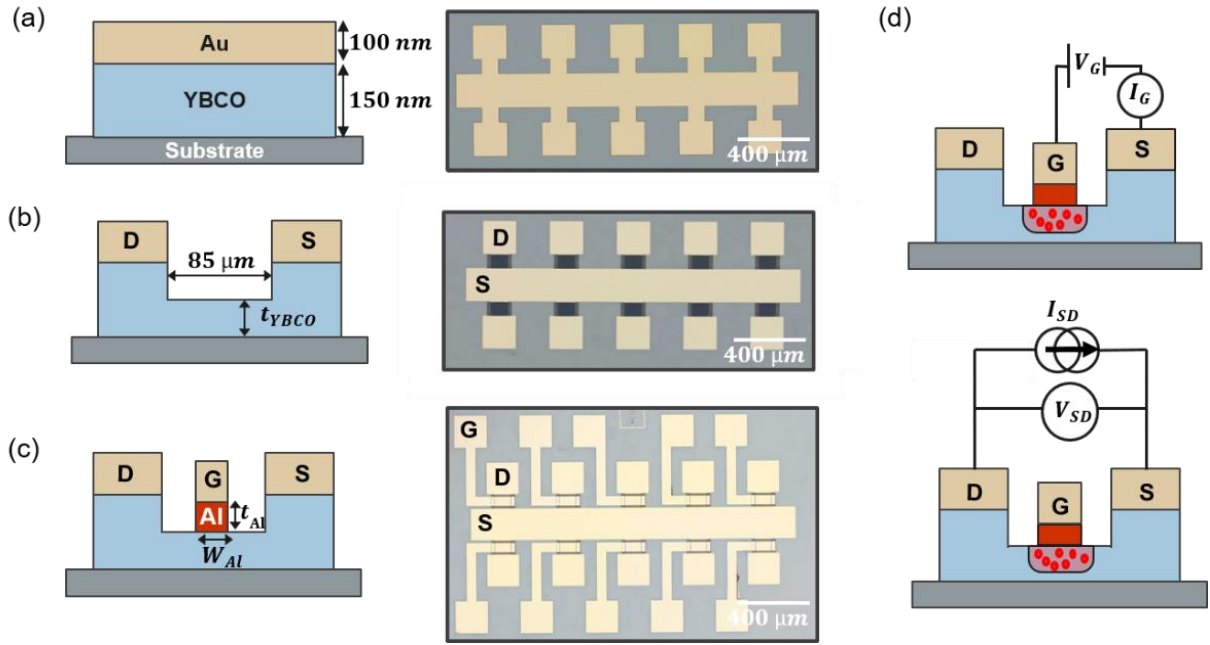

**FIG. 1**. Fabrication process and electrical measuring setup. (a) Etching of the general layout of the device down to the substrate. (b) Etching of the superconducting channel with a thickness $35\ nm < t_{YBCO} < 140\ nm$ while defining the source (S) and drain (D) contacts. The darker areas on the microscope picture correspond to the upper part of the YBCO channel. (c) Deposition step of the Al and Au gate (G) contacts with $t_{Al} = 2\ nm\ or\ 10\ nm$ and $W_{Al} = 5\ \mu m\ or\ 60\ \mu m$. (d) Schematic representation of the measuring setup used for the transport experiments. The electrical characterization is divided into two distinct sets: (top) the gate measurements are carried by applying a voltage bias $V_G$ and measuring the current $I_G$ between G and S and (bottom) the source-drain measurements are carried by injecting a current $I_{SD}$ and measuring the voltage $V_{SD}$ between D and S.

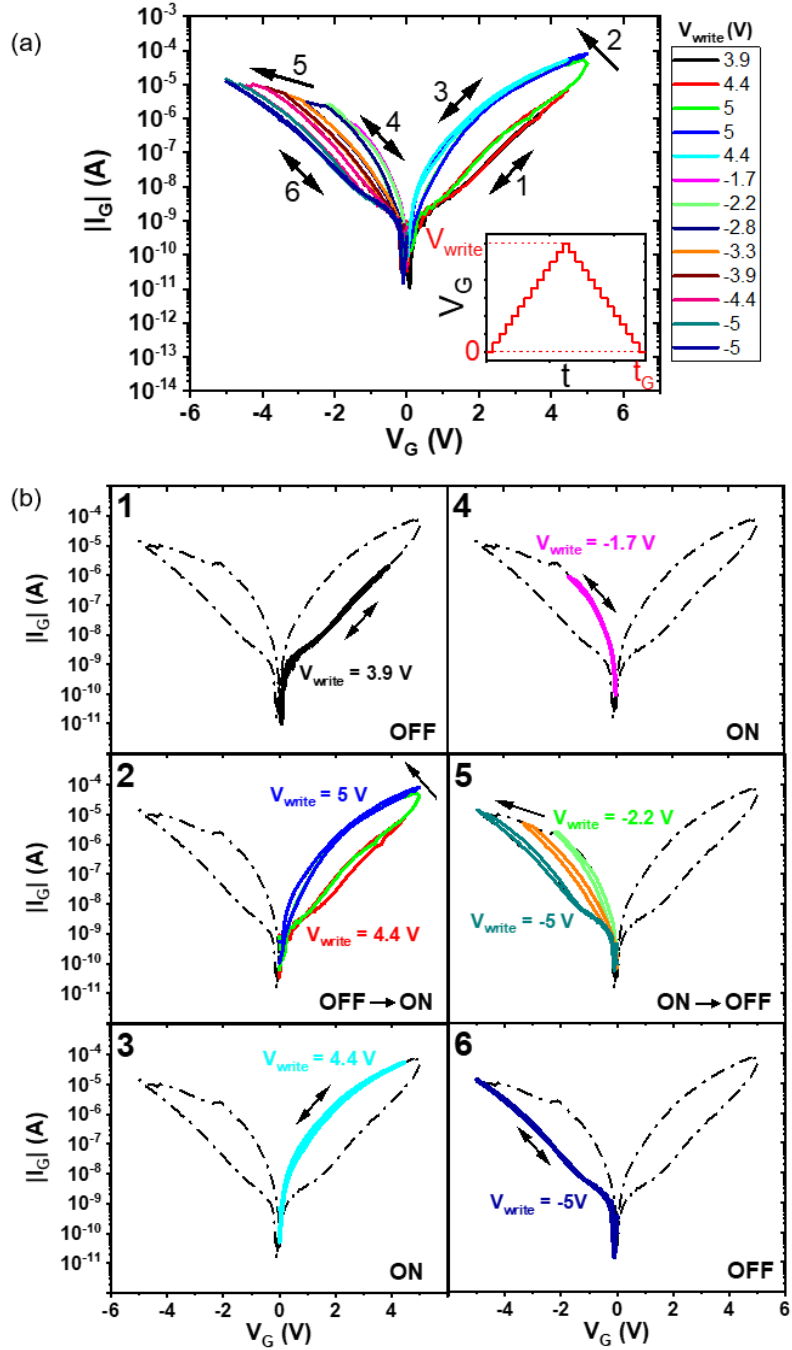

**FIG. 2.** (a) Semilog $I_G$-$V_G$ characteristics of the gate between $-5\,V$ and $+5\,V$ for a device with $t_{YBCO} = 110\,nm$ and $t_{Al} = 10\,nm$ at $40\,K$. The inset shows the shape of the $V_G$ sweep going from $0\,V$ to $V_{write}$ (positive or negative) and back to $0\,V$ in a time $t_G = 20\,s$ and number of steps $n = 100$. (b) Splitting of the switching process into 6 consecutive regimes. For clarity, the contour of the set of I-V characteristics is plotted as a guide to the eye along with selected curves. Graph 2 demonstrates the switching from the low conducting OFF state to the high conducting ON state for gate voltages between $4.4\,V$ and $5\,V$, whereas graph 5 shows the opposite switching as negative voltages are applied to the gate.

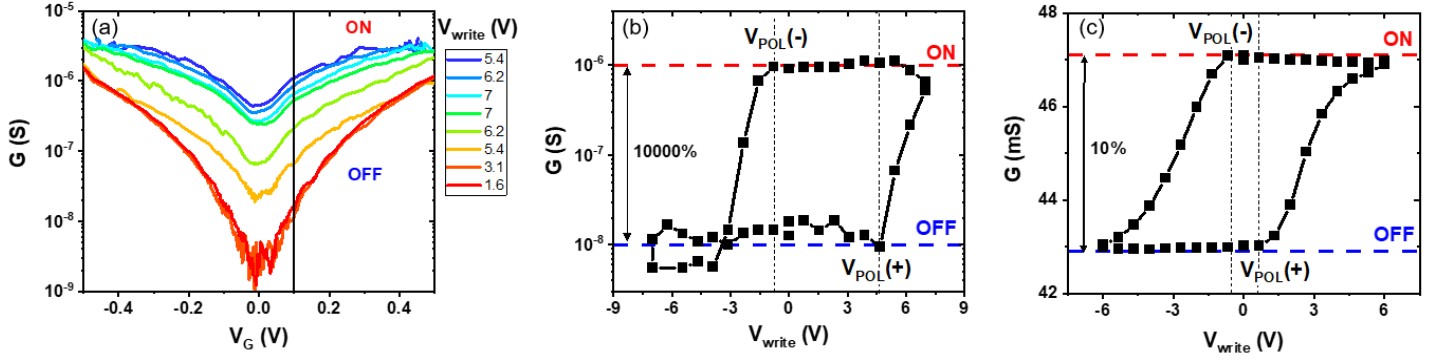

**FIG. 3**. (a) Differential conductance G measured between G and S in semilog scale as a function of a low reading voltage $V_G$ between $-0.5\ V$ and $0.5\ V$ showing a clear switching from the OFF state (blue curve) to the ON state (red curve) as well as intermediate stable states for a device with $t_{YBCO} = 50\ nm$ and $t_{Al} = 10\ nm$ at $35\ K$. $V_{write}$ is applied sequentially as indicated by the legend, from bottom to top. The vertical black line represents the chosen value of $V_G = 100\ mV$ along which **Fig. 3(b)** is calculated. (b) G as a function of $V_{write}$ for a fixed reading voltage of $100\ mV$ demonstrating the reversible switching (with an amplitude of $10000\ \%$) behaviour of G between two non-volatile ON and OFF states. The polarization voltages $V_{POL}(+)$ and $V_{POL}(-)$ represent the minimum value of the voltage required to induce a switching from the OFF to the ON state and inversely, respectively. (c) G as a function of $V_{write}$ for a control sample with $t_{YBCO} = 72\ nm$ at $90\ K$.

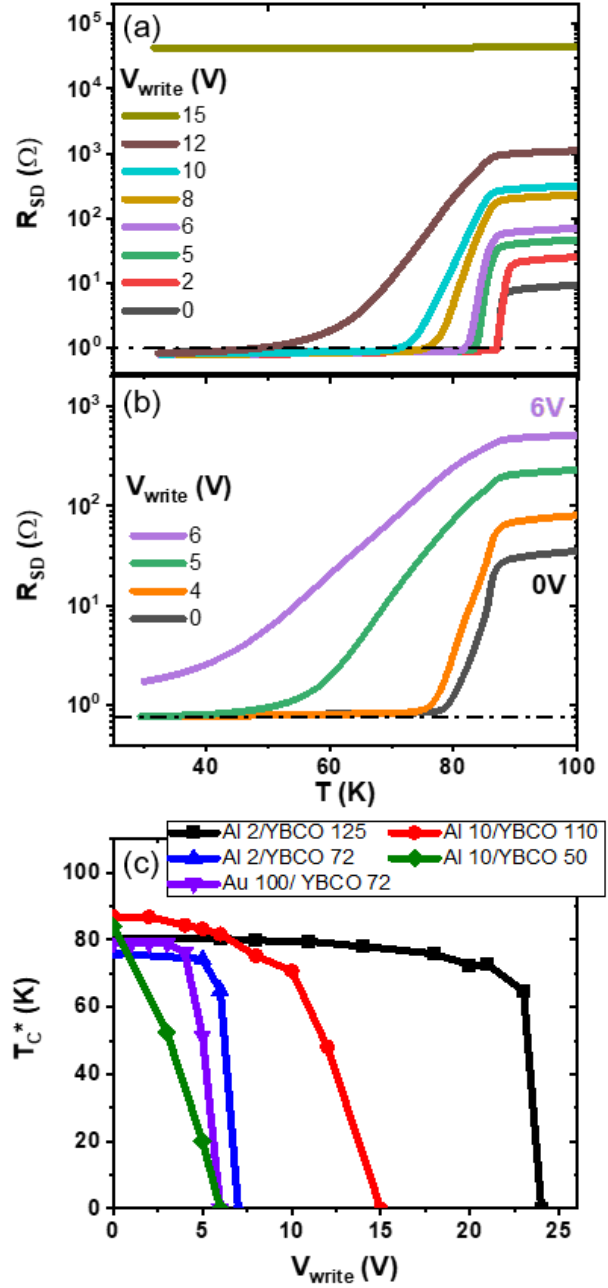

**FIG. 4**. (a) Temperature dependence of the resistance $R_{SD}$ in semilog scale measured after applying $V_{write}$ (see legend) for a device with $t_{YBCO} = 110\ nm$ and $t_{Al} = 10\ nm$. Notice that the resistance drops to a finite value of ~0.9 Ω instead of zero because the two-probe configuration measures the wiring and contact resistances. (b) Temperature dependence of $R_{SD}$ in semilog scale for a control device with no aluminium gating and $t_{YBCO} = 72\ nm$, measured after applying $V_{write}$ (see legend). (c) Effective critical temperature $T_C^*$ of the superconducting channel as a function of $V_{write}$ for different devices with a channel thickness varying between $50\ nm < t_{YBCO} < 125\ nm$.

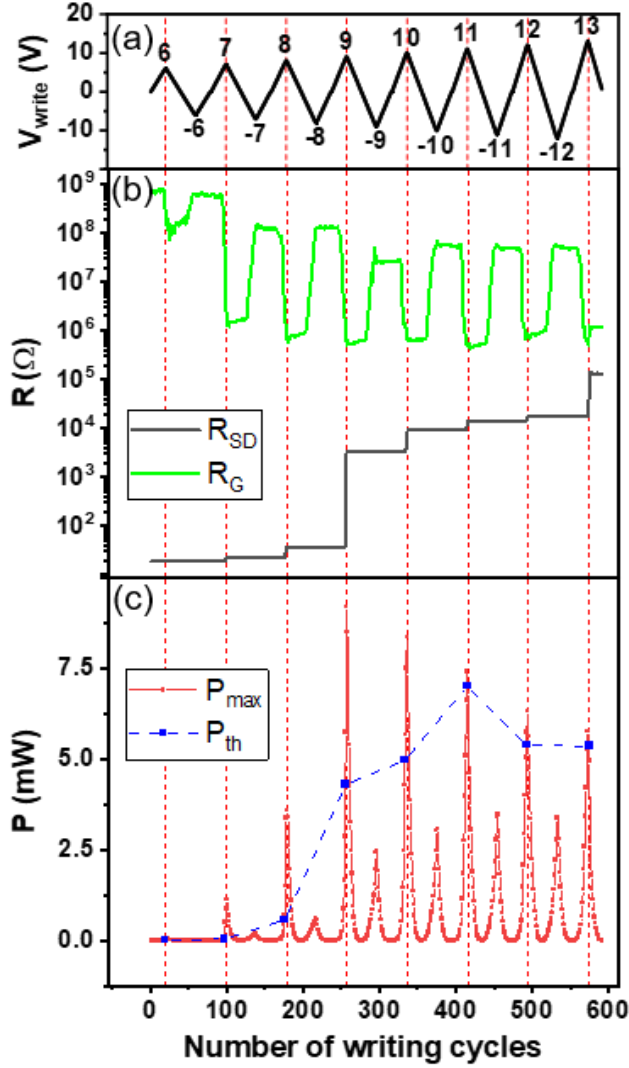

**FIG. 5.** Measurements done at 87 $K$ on a transistor with $t_{YBCO} = 125\ nm$ and $t_{Al} = 10\ nm$. (a) Sequenc of applied $V_{write}$ as a function of the number of writing cycles. (b) Evolution of $R_{SD}$ and $R_G$ as a function of the number of writing cycles. $R_G$ reversibly switches between two resistive levels while $R_G$ increases from 20 $\Omega$ to 0.1 $M\Omega$. (c) Maximum dissipated power P as a function of the number of writing cycles. The highest peaks correspond to the power measured while applying positive voltages (see vertical red dotted lines), whereas the small peaks correspond to the power measured while applying negative voltages. The blue dots represent the values of the threshold power $P_{th}$ required to trigger the non-reversible increase of $R_{SD}$.